\documentclass[a4paper]{article}
\usepackage{amsmath}
\usepackage{latexsym}
\usepackage{graphicx}
\usepackage{mathptmx}
\usepackage{bm}

\title{Extension the Noether's theorem to Lagrangian formulation with nonlocality}

\author{Zaixing Huang\\[5pt]
  State Key Laboratory of Mechanics and Control of Mechanical Structures\\
  Nanjing University of Aeronautics and Astronautics\\
  Yudao Street 29, Nanjing, 210016, P R China\\ E-mail: huangzx@nuaa.edu.cn}

\begin{document}

\maketitle

\begin{abstract}
A Lagrangian formulation with nonlocality is investigated in this
paper. The nonlocality of the Lagrangian is introduced by a new
nonlocal argument that is defined as a nonlocal residual satisfying
the zero mean condition. The nonlocal Euler-Lagrangian equation is
derived from the Hamilton's principle. The Noether's theorem is
extended to this Lagrangian formulation with nonlocality. With the
help of the extended Noether's theorem, the conservation laws
relevant to energy, linear momentum, angular momentum and the
Eshelby tensor are determined in the nonlocal elasticity associated
with the mechanically based constitutive model. The results show
that the conservation laws exist only in the form of the integral
over the whole domain occupied by body. The localization of the
conservation laws is discussed in detail. We demonstrate that not
every conservation law corresponds to a local equilibrium equation.
Only when the nonlocal residual of conservation current exists, can
a conservation law be transformed into a local
equilibrium equation by localization.
\end{abstract}

\section{Introduction} \label{intro} Nonlocal elasticity has
developed into an important branch of continuum mechanics since the
first pioneer studies in the last 60--70s \cite{1,2,3,4,5}. This
theory and the extension of it have been applied to various topics
in engineering \cite{5,6,7,8}. So far, constitutive models in the
nonlocal elasticity can be categorized into three sorts : Eringen's
constitutive model \cite{4,7}, peridynamic constitutive model
\cite{9,10,11} and mechanically based constitutive model
\cite{12,13}. The postulation of the Eringen' constitutive model
consists in that the stress at a point not only depends on strain of
the point but also on strains of all points within body. So stress
is equal to the convolution integral of strain over the domain
occupied by the body. Due to this fact, the mixed boundary value
problem is ill-posed in the nonlocal elasticity associated with the
Eringen's constitutive model \cite{14}, except the attenuating
kernel being equipped some ad hoc futures, for example, it is
assumed to be the Green function of the differential operator
$\nabla^2$. Under the latter assumption, the integral-type
constitutive model reduces to the so-called implicit gradient
model.\\

In the peridynamic constitutive model \cite{9,10,11}, ones introduce
the internal long-range body force to represent the interactions
within body, but forsake the conception of stress and strain. An
integral operator is used to formulate the constitutive relation
between the internal long-range body force and displacements. Since
there are no assumptions made on the differentiability of
displacements in the motion equation of the peridynamic nonlocal
elasticity (peridynamics), this theory is suitable to study
phenomena with discontinuities and fragmentation. In the
peridynamics, no boundary conditions appear as there are no spatial
derivatives. In order to solve the boundary value problems, some
unconventional boundary conditions have been prescribed
\cite{11,15}. Unlike the conventional boundary conditions, they are
imposed on a boundary layer with non-zero volumetric measure, as
opposed to a geometric boundary in the strict mathematical meanings.
This inconsistency may give rise to some difficulties in the
problems with complicated
boundary conditions \cite{12,13}.\\

The mechanically based constitutive model (MBCM) can be regarded as
a fusion between the Eringen's constitutive model and the
peridynamic constitutive model \cite{16}. This model retains the
conception of stress and strain but introduces meanwhile the
internal long-range body force characterizing the interactions
between non-adjacent particles. The internal long-range body force
linearly or nonlinearly depends on the relative displacements
between interacting particles within body; while the stress-strain
relation is still characterized by the conventional constitutive
equation, e.g., the Hooke's law. Stress and the internal long-range
body force are independent on each other. If a nonlocal elasticity
is concerned with MBCM, we refer it as to the nonlocal elasticity
associated with MBCM. In this theory, all types of boundary
conditions are the same as that in the classical elasticity. From
the view of this point, it is of advantage to adopt MBCM in
nonloacal elasticity.\\

Paola et al \cite{12,13} firstly advanced the linear theory of the
nonlocal elasticity associated with MBCM. They called it the
mechanically based model of nonlocal elasticity. In this theory, the
internal long-range body force at a particle is linearly dependent
of the relative displacements between the particle and other
particles within body. If taking a suitable nonlocal kernel, this
internal long-range body force can be also obtained by linearization
of the the peridynamic constitutive model.\\

Recently, the Lagrangian formulation has been proposed for the
mechanically based model of nonlocal elasticity \cite{17}. The
relevant energy-momentum tensor is given under a simplest case in
which the nonlocal kernel degrades into a constant equal to the
reciprocal of the volume of body. The energy-momentum tensor is of
fundamental importance because it represents the configurational
force on a defect (e.g., vacancy, inclusion, dislocation and crack
etc) in solids; while in the absence of defects, it embodies a
conservation law \cite{18}. In order to determine, under a more
general case, the energy-momentum tensor and the relevant
conservation laws, it is necessary to extend the Noether's theorem
to the nonlocal elasticity associated with MBCM.\\

The study on the nonlocal form of the Noether's theorem can be
traced back to Edelen \cite{19}. He \cite{19,20,21} investigated the
reformulation of the Noether theorem in the nonlocal theory based on
a general theoretical framework. In this framework, the nonlocal
argument specified by Edelen \cite{19} is too general in form to
take account of the constraint of physical laws. Later, Edelen
\cite{22} simplified the nonlocal argument as a linear integral
operator on the field variable. Under this case, Vukobrat and
Kuzmannovic \cite{23}) addressed the conservation laws in the
nonlocal elasticity associated with the Eringen's constitutive
model. Recently, Lazar and Kirchner \cite{24} discussed the
energy-momentum tensor and relevant configurational forces in the
nonlocal theory. They issued some interesting results on the
interaction between dislocation and disclination.\\

At present, few results are known on the form of the Noether's
theorem and relevant conservation laws in the nonlocal elasticity
associated with MBCM. So the objective of current work is to clarify
these subjects. The paper is organized as follows: we start in
Section 2 by using a nonlinear integral operator to define the
nonlocal argument. Based on this nonlocal argument, we develop a
Lagrange formulation with nonlocality. The Noether's theorem is
extended into the new Lagrangian formulation in Section 3. According
to the extended the Noether's theorem, in Section 4 we investigate
the conservation laws in the nonlocal elasticity associated MBCM in
which the internal long-range body force is nonlinearly dependent of
the relative displacements between particles. In section 5, the
localization of the conservation laws is discussed. The relevant
nonlocal residuals are determined. We close this paper
in Section 6 by making some concluding remarks.\\

 \textbf{Notation:} A compact notation is used, with boldface
 letters being vectors or tensors. The index rules and
summation convention are adopted. Latin indices have range 1, 2, 3;
while Greek indices run from 0 to 3. Partial derivatives with
respect to coordinates are represented as $a_{,k}=\partial
a/\partial x^k$. Partial derivative with respect to time is denoted
by an upper dot, i.e., $\dot{a}=\partial a/\partial t$. The norm
$|\cdot|$ is defined as $\sqrt{(\cdot)(\cdot)}$. For example,
$|r|=\sqrt{r\cdot r}$. Other symbols will be introduced in the text
where they appear for the first time.
\section{Lagrangian formulation with nonlocality}
\label{sec:1} A continuum occupies the domain $\Omega $ in the
three-dimensional Euclidean space. Let every particle in the
continuum be referred to by the orthogonal Cartesian coordinates
$\textbf{x}=\{x^1, x^2, x^3\}$ specifying its position in $\Omega $,
and let $\varphi=\varphi(t,\textbf{x})$ denote a field variable
defined on $\Omega $. Depending on circumstances, $\varphi$ is a
scalar, vector or tensor. Let $\langle h|\varphi\rangle$ represent
the nonlocal argument of $\varphi$. We define it as
\begin{equation}\label{1}\langle h|\varphi\rangle=
\varphi(t,\textbf{x})\int_\Omega h(\textbf{x}, \textbf{y},
|\varphi(t,\textbf{x})-\varphi(t,\textbf{y})|)
\mathrm{d}v(\textbf{y}) -\int_\Omega h(\textbf{x}, \textbf{y},
|\varphi(t,\textbf{x})-\varphi(t,\textbf{y})|)\varphi(t,\textbf{y})\mathrm{d}v(\textbf{y}),\end{equation}        
where
$h(\textbf{x},\textbf{y},|\varphi(t,\textbf{x})-\varphi(t,\textbf{y})|)$
is called the nonlocal kernel. Let
$r=\varphi(t,\textbf{x})-\varphi(t,\textbf{y})$. Eq. (\ref{1}) can
be abbreviated to
\begin{eqnarray}\label{1s}\langle h|\varphi\rangle&=&\varphi(t,\textbf{x})\int_\Omega h(\textbf{x}, \textbf{y},
|r|) \mathrm{d}v(\textbf{y}) -\int_\Omega h(\textbf{x}, \textbf{y},
|r|)\varphi(t,\textbf{y})\mathrm{d}v(\textbf{y})\nonumber\\&=&\int_\Omega
h(\textbf{x}, \textbf{y},
|r|)[\varphi(t,\textbf{x})-\varphi(t,\textbf{y})]\mathrm{d}v(\textbf{y})\nonumber\\&=&
\int_\Omega h(\textbf{x}, \textbf{y}, \mid r \mid)
r\mathrm{d}v(\textbf{y}).\end{eqnarray}                                        
Different from the case recently reported by Huang \cite{17}, the
nonlocal kernel depends not only on \textbf{x} and \textbf{y} but
also on the field variables. Therefore, $\langle h|\varphi\rangle$
is a nonlinear integral operator with respect to $\varphi$. We
enforce the nonlocal kernel to fulfil the symmetry: $h(\textbf{x},
\textbf{y},|r|)=h(\textbf{y}, \textbf{x},|r|)$. Under this symmetry,
it is easy to verify that $\langle h|\varphi\rangle$ satisfies the
zero mean condition:
\begin{equation}\label{2}\int_\Omega
\langle h|\varphi\rangle\mathrm{d}v(\textbf{x})=0.\end{equation}    
Due to Eq.(\ref{2}), the nonlocal argument determined by Eq.
(\ref{1}) is in essence different from the definition given by
Edelen \cite{19,22}. The latter fails to satisfy the zero mean
condition.\\

Let $L=L(t, \textbf{x}, \varphi, \dot{\varphi}, \varphi_{,k},
\langle h|\varphi\rangle) \, (k=1,2,3)$ denote the Lagrangian
\footnote[1]{If necessary, the nonlocal arguments $\langle
h|\dot{\varphi}\rangle$ and $\langle h|\varphi_{,k}\rangle$ may be
conveniently inserted into \emph{L}. But in this case, the boundary
conditions will become too complicated to solve in mathematics.}. So
the action functional of $\varphi$ can be written as
\begin{equation}\label{3}A[\varphi]=\int_{t_0}^{t_1}\int_\Omega L(t, \textbf{x},
\varphi,\dot{\varphi},\varphi_{,k},
\langle h|\varphi\rangle)\mathrm{d}v(\textbf{x})\mathrm{d}t.\end{equation}                
Suppose $\varphi(t,\textbf{x})$, $h(\textbf{x}, \textbf{y},|r|)$ and
$L$ are suitably smooth functions. In order to determine the
variation $\delta A[\varphi]$, we firstly prove the two identities.
Using Eq. (\ref{1s}), we have
\begin{eqnarray}\label{r1}\delta \langle h|\varphi\rangle&=&
\int_\Omega\{[\delta h(\textbf{x}, \textbf{y}, |r|)]r+h(\textbf{x},
\textbf{y}, |r|)\delta r\}\mathrm{d}v(\textbf{y})=\int_\Omega
(h+|r|\frac{\partial
h}{\partial |r|})\delta r\mathrm{d}v(\textbf{y})\nonumber\\
&=&\delta\varphi(t,\textbf{x})\int_\Omega (h+|r|\frac{\partial
h}{\partial |r|}) \mathrm{d}v(\textbf{y}) -\int_\Omega
(h+|r|\frac{\partial
h}{\partial |r|})\delta\varphi(t,\textbf{y})\mathrm{d}v(\textbf{y})\nonumber\\
&=&\langle g|\delta\varphi\rangle.\end{eqnarray}                                   
\begin{eqnarray}\label{r2}\int_\Omega\psi (t,\textbf{x})\langle
h|\varphi\rangle\mathrm{d}v(\textbf{x})&=&\int_\Omega\int_\Omega\psi
(t,\textbf{x})h(\textbf{x},\textbf{y},|r|)[\varphi(t,\textbf{x})-\varphi(t,\textbf{y})]\mathrm{d}v(\textbf{y})\mathrm{d}v(\textbf{x})\nonumber\\
&=&\int_\Omega\int_\Omega\varphi
(t,\textbf{x})h(\textbf{x},\textbf{y},|r|)[\psi(t,\textbf{x})-\psi(t,\textbf{y})]\mathrm{d}v(\textbf{y})\mathrm{d}v(\textbf{x})\nonumber\\
&=&\int_\Omega\varphi (t,\textbf{x})\langle
h|\psi\rangle\mathrm{d}v(\textbf{x}).\end{eqnarray}                   
Eq. (\ref{r2}) is valid for any continuous function $h$, $\psi$ and
$\varphi$. For example, we may use $g$ in Eq. (\ref{r1}) to replace
$h$, where $g$ reads
\begin{equation}\label{4}g=h+\mid
r\mid\frac{\partial h}{\partial \mid r\mid}=h(\textbf{x},
\textbf{y}, \mid r\mid)+\mid r\mid\frac{\partial h(\textbf{x},
\textbf{y}, \mid r\mid)}{\partial
\mid r\mid}.\end{equation}\\                                                     
Clearly, $g$ is symmetric with respect to \textbf{x} and \textbf{y}.
By means of Eq. (\ref{r1}) and Eq.(\ref{r2}), we can calculate the
variation of $A[\varphi]$:
\begin{eqnarray}\label{5}\delta A[\varphi]
&=&\int_{t_0}^{t_1}\int_\Omega(\frac {\partial
L}{\partial\varphi}\delta\varphi+\frac{\partial
L}{\partial\dot{\varphi}}\delta\dot{\varphi}+\frac{\partial
L}{\partial\varphi_{,k}}\delta\varphi_{,k}+\frac{\partial
L}{\partial \langle h|\varphi\rangle}\delta\langle h|\varphi\rangle)
\mathrm{d}v(\textbf{x})\mathrm{d}t\nonumber\\
&=&\int_{t_0}^{t_1}\int_\Omega[\frac{\partial
L}{\partial\varphi}-\frac{\mathrm{d}}{\mathrm{d}t}(\frac {\partial
L}{\partial\dot{\varphi}})-(\frac{\partial
L}{\partial\varphi_{,k}})_{,k}+\langle g|\frac{\partial L}{\partial
\langle h|\varphi\rangle}\rangle]\delta\varphi
\mathrm{d}v(\textbf{x})\mathrm{d}t\nonumber\\
&+&\int_\Omega\left. \frac {\partial
L}{\partial\dot{\varphi}}\delta\varphi\right|^{t_1}_{t_0}\mathrm{d}v(\textbf{x})
+\int_{t_0}^{t_1}\int_{\partial\Omega}\frac{\partial
L}{\partial\varphi_{,k}}n_k\delta\varphi
\mathrm{d}s(\textbf{x})\mathrm{d}t,                                            
\end{eqnarray}
where $\partial\Omega$ is the boundary surface of $\Omega$ and $n_k$
denotes the unit normal vector on $\partial\Omega$. In Eq.
(\ref{5}), a shortened form similar to Eq.(\ref{1}) is used,
\begin{equation}\label{6}\langle g|\frac{\partial L}{\partial
\langle h|\varphi\rangle}\rangle=\frac{\partial L}{\partial \langle
h|\varphi\rangle}\int_\Omega\ g(\textbf{x}, \textbf{y}, \mid
r\mid)\mathrm{d}v(\textbf{y})- \int_\Omega\ g(\textbf{x},
\textbf{y}, \mid r\mid)\frac{\partial L}{\partial \langle
h|\varphi\rangle}
\mathrm{d}v(\textbf{y}).\end{equation}\\                                           

Let $\partial\Omega=\partial\Omega_1\cup\partial\Omega_2$,
$\partial\Omega_1\cap\partial\Omega_2=\emptyset$. On
$\partial\Omega_1$, $\varphi$ takes a given value $\bar\varphi$.
Then, the boundary condition on $\partial\Omega_1$ reads
\begin{equation}\label{7}\left. \varphi\right|_{\partial\Omega_1}
=\bar\varphi.\end{equation}                                         
At the initial and terminal time, we have
\begin{equation}\label{8}\left.
\varphi\right|_{t_0}=\bar\varphi_0,
 \quad \left. \varphi\right|_{t_1}=\bar\varphi_1.\end{equation}                    
Due to Eq.(\ref{7}), $\delta\varphi=0$ on $\partial\Omega_1$.
Similarly, $\delta\varphi=0$ at the initial and terminal time
because of Eq.(\ref{8}). Thus, Eq.(\ref{5}) reduces to
\begin{equation}\label{9}\delta A[\varphi]=\int_{t_0}^{t_1}\int_\Omega[\frac {\partial
L}{\partial\varphi}-\frac{\mathrm{d}}{\mathrm{d}t}(\frac {\partial
L}{\partial\dot{\varphi}})-(\frac{\partial
L}{\partial\varphi_{,k}})_{,k}+\langle g|\frac{\partial L}{\partial
\langle h|\varphi\rangle}\rangle]\delta\varphi
\mathrm{d}v(\textbf{x})\mathrm{d}t+\int_{t_0}^{t_1}\int_{\partial\Omega_2}\frac{\partial
L}{\partial\varphi_{,k}}n_k\delta\varphi
\mathrm{d}s(\textbf{x})\mathrm{d}t.                                                                
\end{equation}
In terms of the Hamilton's principle, we have $\delta A[\varphi]=0$.
So the fundamental lemma of
variation yields the below results:\\
Euler-Lagrange equation:
\begin{equation}\label{10}\frac{\mathrm{d}}{\mathrm{d}t}(\frac {\partial
L}{\partial\dot{\varphi}})+(\frac{\partial
L}{\partial\varphi_{,k}})_{,k}-\frac {\partial
L}{\partial\varphi}=\langle g|\frac{\partial L}{\partial
\langle h|\varphi\rangle}\rangle.\end{equation}                         
Natural boundary condition:
\begin{equation}\label{11}\left. \frac{\partial L}{\partial\varphi_{,k}}n_k
\right|_{\partial\Omega_2}=0.\end{equation}                                        
Eq.(\ref{10}) is also referred to as the nonlocal Euler-Lagrange
equation. The right-side term of Eq.(\ref{10}) is the nonlocal term,
called the nonlocal traction. If $h$ is independent of $r$, we have
$g=h$, and then Eq.(\ref{10}) will reduce to the case in \cite{17}.\\

Using the symmetry of $g(\textbf{x}, \textbf{y}, |r|)$, we easily
verify the equality below:
\begin{equation}\label{12}\int_\Omega
\frac{\partial L(\textbf{x})}{\partial \langle
h|\varphi\rangle}\int_\Omega g(\textbf{x}, \textbf{y}, \mid
r\mid)\mathrm{d}v(\textbf{y})\mathrm{d}v(\textbf{x})=\int_\Omega\int_\Omega
g(\textbf{x}, \textbf{y}, \mid r\mid)\frac{\partial
L(\textbf{y})}{\partial
\langle h|\varphi\rangle}\mathrm{d}v(\textbf{y})\mathrm{d}v(\textbf{x}),\end{equation}  
by interchanging \textbf{x} and \textbf{y}. As thus, the integral of
Eq.(\ref{6}) over $\Omega$ leads to
\begin{equation}\label{13}\int_\Omega
\langle g|\frac{\partial L}{\partial
\langle h|\varphi\rangle}\rangle\mathrm{d}v(\textbf{x})=0,\end{equation}                       
which shows that the nonlocal traction is a nonlocal residual
automatically satisfying the zero mean condition. Due to
Eq.(\ref{13}), the integral of Eq.(\ref{10}) over $\Omega $ has the
same expression as the integral of the
Euler-Lagrangian equation not concerned with nonlocality.\\

In physics, if Eq. (\ref{10}) represents the equation of motion, the
nonlocal traction may be interpreted as an internal long-range body
force applied on a particle at \textbf{x} by other particles within
body. Every particle is subjected to such a force. In terms of the
action and reaction law, the sum of all such forces must be zero.
Therefore, Eq.(\ref{13}) is just an embodiment of the action and
reaction law.
\section{Extension of the Noether's theorem to the Lagrangian formulation with nonlocality}
\label{sec:2} For convenience, time can be treated as a coordinate
so as to form a four-dimensional position vector $\hat{\bf{x}}$ with
coordinates $\hat{\bf{x}}=\{x^\beta\}=\{x^0, x^1, x^2, x^3\}$
defined on $\hat\Omega$, where $x^0=t$ and
$\hat\Omega=[t^0,t^1]\cup\Omega$. Thus, the Lagrangian function can
be abbreviated as $L=L(\hat{\textbf{x}}, \varphi,\varphi_{,\gamma},
\langle\varphi\rangle), \, (\gamma=0,1,2,3)$, and the nonlocal
Euler-lagrange equation (see Eq.(\ref{10})) can be rewritten as
\begin{equation}\label{14}(\frac{\partial
L}{\partial\varphi_{,\gamma}})_{,\gamma}-\frac {\partial
L}{\partial\varphi}=\langle g|\frac{\partial L}{\partial
\langle h|\varphi\rangle}\rangle.\end{equation}              
Consider infinitesimal transformations of group
\begin{eqnarray}\label{15}
\hat{\bf{x}}&\longmapsto&\hat{\bf{x}}'=\hat{\bf{x}}+\delta \hat{\bf{x}}.\\
\varphi(\hat{\bf{x}})&\longmapsto&\bar{\varphi}'(\hat{\bf{x}}')=\varphi(\hat{\bf{x}})+\Delta\varphi(\hat{\bf{x}}).\\
\hat{\bf{y}}&\longmapsto&\hat{\bf{y}}'=\hat{\bf{y}}+\delta \hat{\bf{y}}.\\
\varphi(\hat{\bf{y}})&\longmapsto&\bar{\varphi}'(\hat{\bf{y}}')=\varphi(\hat{\bf{y}})+\Delta\varphi(\hat{\bf{y}}).
\end{eqnarray}                                                           
Here, $\Delta\varphi$ is different from the variation
$\delta\varphi$. The latter is defined as
\begin{equation}\label{19}\delta\varphi=\varphi'(\hat{\bf{x}})-\varphi(\hat{\bf{x}}).\end{equation}          
Therefore, between $\Delta\varphi$ and $\delta\varphi$, there exists
the relation below:
\begin{equation}\label{20}\Delta\varphi=\delta\varphi+\varphi_{,\gamma}\delta x^\gamma.\end{equation}          
The action functional $A[\varphi]$ is said to be symmetry if it is
form-invariant with respect to the infinitesimal transformations
(\ref{15}) -- (21), i.e.,
\begin{equation}\label{21}\int_{\hat\Omega'}L[\hat{\bf{x}}',\varphi'(\hat{\bf{x}}'),
\varphi'_{,\gamma}(\hat{\bf{x}}'),\langle
h|\varphi'\rangle]\mathrm{d}v(\hat{\bf{x}}')=\int_{\hat\Omega}L[\hat{\bf{x}},\varphi(\hat{\bf{x}}),
\varphi_{,\gamma}(\hat{\bf{x}}),\langle
h|\varphi\rangle]\mathrm{d}v(\hat{\bf{x}}).\end{equation}                    
In the left integral, $\hat{\bf{x}}'$ now represents merely a dummy
variable of integration and can therefore be relabeled
$\hat{\bf{x}}$. But there remains a change in the domain of
integration, so Eq. (\ref{21}) becomes
\begin{equation}\label{22}\int_{\hat\Omega'}L[\hat{\bf{x}},\varphi'(\hat{\bf{x}}),
\varphi'_{,\gamma}(\hat{\bf{x}}),\langle
h|\varphi'\rangle]\mathrm{d}v(\hat{\bf{x}})=\int_{\hat\Omega}L[\hat{\bf{x}},\varphi(\hat{\bf{x}}),
\varphi_{,\gamma}(\hat{\bf{x}}),\langle
h|\varphi\rangle]\mathrm{d}v(\hat{\bf{x}}).\end{equation}                    
It should be noted that $\langle h|\varphi'\rangle$ in the left-hand
term of Eq. (\ref{22}) is an integral defined on $\Omega'$, but the
integral domain of $\langle h|\varphi\rangle$ in the right-hand term
is $\Omega$. For $\langle h|\varphi'\rangle$, we have
\begin{equation}\label{23}\langle h|\varphi'\rangle=
\varphi'(\hat{\bf{x}})\int_{\Omega'} h({\bf{x}}, {\bf{y}}, |r'|)
\mathrm{d}v({\bf{y}}) -\int_{\Omega'} h({\bf{x}}, {\bf{y}},
|r'|)\varphi'(\hat{\bf{y}})\mathrm{d}v({\bf{y}}).\end{equation}        
By the transport theorem \cite{25}, Eq. (\ref{23}) leads to
\begin{eqnarray}\label{24}\langle h|\varphi'\rangle&=&
\varphi'(\hat{\bf{x}})\{\int_{\Omega} h({\bf{x}}, {\bf{y}}, |r'|)
\mathrm{d}v({\bf{y}})+\int_{\Omega} \frac{\partial[h({\bf{x}},
{\bf{y}}, |r'|)\delta y^k]}{\partial y^k} \mathrm{d}v({\bf{y}})\}\nonumber\\
&-&\int_{\Omega} h({\bf{x}}, {\bf{y}},
|r'|)\varphi'(\hat{\bf{y}})\mathrm{d}v({\bf{y}})-\int_{\Omega}
\frac{\partial[h({\bf{x}}, {\bf{y}},
|r'|)\varphi'(\hat{\bf{y}})\delta y^k]}
{\partial y^k} \mathrm{d}v({\bf{y}}).\end{eqnarray}                              
Expanding $h({\bf{x}}, {\bf{y}}, |r'|)$ at $r$ to the first-order
term yields
\begin{eqnarray}\label{25}
h({\bf{x}}, {\bf{y}}, |r'|)&=&h({\bf{x}}, {\bf{y}},
|r|)+\frac{\partial h}{\partial |r|}\frac{r}{|r|}\delta
r\nonumber\\&=&h({\bf{x}}, {\bf{y}}, |r|)+\frac{\partial h}{\partial
|r|}\frac{\varphi(\hat{\bf{x}})-\varphi(\hat{\bf{y}})}{|r|}[\delta\varphi(\hat{\bf{x}})
-\delta\varphi(\hat{\bf{y}})].\end{eqnarray}                                                    
Inserting Eq. (\ref{19}) and (\ref{25}) in (\ref{24}) and omitting
the higher-order terms, we have
\begin{equation}\label{26}\langle
h|\varphi'\rangle=\langle h|\varphi\rangle+\langle
g|\delta\varphi\rangle+\int_{\Omega} \frac{\partial[h({\bf{x}},
{\bf{y}}, |r|) r \delta
y^k]}{\partial y^k} \mathrm{d}v({\bf{y}}).\end{equation}          
After  Eq. (\ref{19}) and (\ref{26})are substituted into
$L[\hat{\bf{x}},\varphi'(\hat{\bf{x}}),
\varphi'_{,\gamma}(\hat{\bf{x}}),\langle h|\varphi'\rangle]$, it
becomes
\begin{eqnarray}\label{27s1}L[\hat{\bf{x}},\varphi'(\hat{\bf{x}}),
\varphi'_{,\gamma}(\hat{\bf{x}}),\langle
h|\varphi'\rangle]&=&L[\hat{\bf{x}},\varphi+\delta\varphi,
\varphi_{,\gamma}+(\delta\varphi)_{,\gamma},\langle
h|\varphi\rangle+\langle g|\delta\varphi\rangle+\int_{\Omega}
\frac{\partial[h({\bf{x}}, {\bf{y}}, |r|) r \delta
y^k]}{\partial y^k} \mathrm{d}v({\bf{y}})]\nonumber\\
&=&L(\hat{\bf{x}},\varphi, \varphi_{,\gamma},\langle
h|\varphi\rangle)+\delta L ,\end{eqnarray}                    
where
\begin{equation}\label{27s2}\delta L=\frac{\partial
L}{\partial\varphi}\delta\varphi+ \frac{\partial
L}{\partial\varphi_{,\gamma}}(\delta\varphi)_{,\gamma}+\frac{\partial
L}{\partial\langle h|\varphi\rangle}[\langle g|\delta\varphi\rangle+
\int_\Omega \frac{\partial[h({\bf{x}}, {\bf{y}}, |r|) r \delta
y^k]}{\partial y^k} \mathrm{d}v({\bf{y}})].\end{equation}                    
By Eq. (\ref{27s1}), we have
\begin{equation}\label{27s3}\int_{\hat\Omega'}L[\hat{\bf{x}},\varphi'(\hat{\bf{x}}),
\varphi'_{,\gamma}(\hat{\bf{x}}),\langle
h|\varphi'\rangle]\mathrm{d}v(\hat{\bf{x}})=\int_{\hat\Omega'}\{L[\hat{\bf{x}},\varphi(\hat{\bf{x}}),
\varphi_{,\gamma}(\hat{\bf{x}}),\langle
h|\varphi\rangle]+\delta L\}\mathrm{d}v(\hat{\bf{x}}).\end{equation}                    
Applying the transport theorem to the right-hand of Eq. (\ref{27s3})
and omitting the higher-order terms, we have
\begin{equation}\label{27s4}\int_{\hat\Omega'}L[\hat{\bf{x}},\varphi'(\hat{\bf{x}}),
\varphi'_{,\gamma}(\hat{\bf{x}}),\langle
h|\varphi'\rangle]\mathrm{d}v(\hat{\bf{x}})=\int_{\hat\Omega}L[\hat{\bf{x}},\varphi(\hat{\bf{x}}),
\varphi_{,\gamma}(\hat{\bf{x}}),\langle h|\varphi\rangle]\mathrm{d}v(\hat{\bf{x}})+
\int_{\hat\Omega}\delta L\mathrm{d}v(\hat{\bf{x}})+
\int_{\hat\Omega}(L\delta x^\gamma)_{,\gamma}.\mathrm{d}v(\hat{\bf{x}}).\end{equation}                    
Substituting Eq. (\ref{27s4}) into (\ref{22}) leads to
\begin{equation}\label{27s5}\int_{\hat\Omega}\delta L\mathrm{d}v(\hat{\bf{x}})+
\int_{\hat\Omega}(L\delta x^\gamma)_{,\gamma}.\mathrm{d}v(\hat{\bf{x}})=0.\end{equation}                    
Using Eq. (\ref{27s2}), we have
\begin{eqnarray}\label{27}\int_{\hat\Omega}\delta L\mathrm{d}v(\hat{\bf{x}})&=&\int_{t_0}^{t_1}\mathrm{d}t\int_{\Omega}\delta L\mathrm{d}v({\bf{x}})
\nonumber\\&=&\int_{t_0}^{t_1}\mathrm{d}t\int_{\Omega}\{\frac{\partial
L}{\partial\varphi}\delta\varphi+ \frac{\partial
L}{\partial\varphi_{,\gamma}}(\delta\varphi)_{,\gamma}+\frac{\partial
L}{\partial\langle h|\varphi\rangle}[\langle g|\delta\varphi\rangle+
\int_{\Omega} \frac{\partial[h({\bf{x}}, {\bf{y}}, |r|) r \delta
y^k]}{\partial y^k}
\mathrm{d}v({\bf{y}})]\}\mathrm{d}v(\hat{\bf{x}})\nonumber\\&=&
\int_{\hat\Omega}\delta\varphi[\frac{\partial
L}{\partial\varphi}-(\frac{\partial
L}{\partial\varphi_{,\gamma}})_{,\gamma}+\langle g|\frac{\partial
L}{\partial\langle
h|\varphi\rangle}\rangle]\mathrm{d}v(\hat{\bf{x}})+\int_{\hat\Omega'}(\frac{\partial
L}{\partial\varphi_{,\gamma}}\delta\varphi)_{,\gamma}\mathrm{d}v(\hat{\bf{x}})\nonumber\\&-&\int_{\hat\Omega}\frac{\partial}{\partial
x^\gamma}[\delta x^\gamma\int_{\Omega} rh({\bf{x}}, {\bf{y}},
|r|)\frac{\partial L}{\partial\langle
h|\varphi\rangle}\mathrm{d}v({\bf{y}})]\mathrm{d}v(\hat{\bf{x}}).\end{eqnarray}                    
By Eq. (\ref{14}), Eq. (\ref{27}) reduces to
\begin{equation}\label{28}\int_{\hat\Omega}\delta L\mathrm{d}v(\hat{\bf{x}})=\int_{\hat\Omega}(\frac{\partial
L}{\partial\varphi_{,\gamma}}\delta\varphi)_{,\gamma}\mathrm{d}v(\hat{\bf{x}})-\int_{\hat\Omega}\frac{\partial}{\partial
x^\gamma}[\delta x^\gamma\int_{\Omega} rh({\bf{x}}, {\bf{y}},
|r|)\frac{\partial L}{\partial\langle
h|\varphi\rangle}\mathrm{d}v({\bf{y}})]\mathrm{d}v(\hat{\bf{x}}).\end{equation}                    
Substituting Eq. ({\ref{28}) into (\ref{27s5}) leads to
\begin{equation}\label{30}\int_{\hat\Omega}[\frac{\partial
L}{\partial\varphi_{,\gamma}}\delta\varphi+L\delta x^\gamma-\delta
x^\gamma\int_{\Omega} rh({\bf{x}}, {\bf{y}}, |r|)\frac{\partial
L}{\partial\langle
h|\varphi\rangle}\mathrm{d}v({\bf{y}})]_{,\gamma}\mathrm{d}v(\hat{\bf{x}})=0.\end{equation}                    
If Eq. (\ref{15}) and (19) belong to a finite Lie group, $\delta
x^\gamma$ and $\Delta\varphi$ can be represented as \cite{26}
\begin{equation}\label{31}\delta x^\gamma=\epsilon^\alpha
X^\gamma_\alpha,\qquad \Delta\varphi=\epsilon^\alpha
\Phi_\alpha,\end{equation}                                                           
where $\epsilon^\alpha$ is an infinitesimal parameter independent of
the space-time coordinates. $X^\gamma_\alpha$ and $\Phi_\alpha$ are
the generators of Lie group. By using Eq. (\ref{31}) and (\ref{20}),
Eq. (\ref{30}) becomes
\begin{equation}\label{32}\int_{\hat\Omega}\{\frac{\partial
L}{\partial\varphi_{,\gamma}}\Phi_\alpha+[(L-\int_{\Omega}
rh({\bf{x}}, {\bf{y}}, |r|)\frac{\partial L}{\partial\langle
h|\varphi\rangle}\mathrm{d}v({\bf{y}}))\delta^\gamma_\mu-
\frac{\partial
L}{\partial\varphi_{,\gamma}}\varphi_{,\mu}]X^\mu_\alpha\}_{,\gamma}
\mathrm{d}v(\hat{\bf{x}})=0.\end{equation}                                               
Thus, the Noether theorem is extended to the nonlocal Lagrangian
formulation with nonlocality. Let
\begin{equation}\label{33}J^\gamma_\alpha=\frac{\partial
L}{\partial\varphi_{,\gamma}}\Phi_\alpha+\{[L-\int_{\Omega}
rh({\bf{x}}, {\bf{y}}, |r|)\frac{\partial L}{\partial\langle
h|\varphi\rangle}\mathrm{d}v({\bf{y}})]\delta^\gamma_\mu-
\frac{\partial L}{\partial\varphi_{,\gamma}}\varphi_{,\mu}\}X^\mu_\alpha,\end{equation}                    
where $J^\gamma_\alpha$ is called the conservation current (or
Noether current). Thus, Eq. (\ref{32}) can be abbreviated as
\begin{equation}\label{34}\int_{\hat\Omega}J^\gamma_{\alpha,\gamma}\mathrm{d}v(\hat{\bf{x}})=0, \qquad \mathrm{or} \qquad
\int_{\partial\hat\Omega}J^\gamma_\alpha n_\gamma\mathrm{d}a(\hat{\bf{x}})=0,\end{equation}                    
where $n_\gamma$ is an unit normal vector on $\partial\hat\Omega$.
Eq.(\ref{34}) shows that total conservation current
$J^\gamma_\alpha$ on $\hat\Omega$ is conserved. However, Eq.
(\ref{34}) will cease to be valid if its integral domain
$\hat\Omega$ is replaced by a subdomain of $\hat\Omega$. This is
because the nonlocal argument is defined on the whole domain
$\Omega$, and it can not be altered in the deductive process from
Eq.(\ref{21}) to (\ref{34}). In other words, if we use any $\hat{v}$
($\hat{v}=[t^0,t^1]\cup v$, $v\subset\Omega$) to replace
$\hat\Omega$ in Eq.(\ref{21}) but keep the integral domain of the
nonlocal argument fixed, it will be impossible to derive Eq.
({\ref{34}) from (\ref{21}) due to needing to change the order of
integrals through the interchange of ${\bf{x}}$ and ${\bf{y}}$.
Therefore, the conservation current is conserved on $\hat\Omega$ as
a whole, and yet when exactly the same statement is made for a
subdomain of $\hat\Omega$ it is no longer valid. As a result, we can
merely derive $J^\gamma_{\alpha,\gamma}=R_\alpha(\textbf{x})$ from
the localization of Eq. (\ref{34}), rather than
$J^\gamma_{\alpha,\gamma}=0$. $R_\alpha(\textbf{x})$ is called the
nonlocal residual of the conservation current. It should satisfy the
zero mean condition so that the integral of
$J^\gamma_{\alpha,\gamma}=R_\alpha(\textbf{x})$ over $\hat\Omega$
can return to Eq. (\ref{34}). Huang put forward a representation of
the nonlocal residual automatically
satisfying the zero mean condition \cite{27}.\\

For the convenience, in subsequent sections we call the Eq.
(\ref{34}) the conservation law, while
$J^\gamma_{\alpha,\gamma}=R_\alpha(\textbf{x})$ is referred to as
the local equilibrium equation.
\section{Conservation laws in nonlocal elasticity}
\label{sec:3} Consider a linear, homogenous, nonlocal elastic body
free of external body forces. For this body, the Lagrangian function
is assumed to take the following form:
\begin{equation}\label{37}L=\frac{1}{2}\rho\dot{u}_i\dot{u}_i-\frac{1}{2}C_{ijkl}u_{i,j}u_{k,l}-
\frac{1}{2}\langle h|u_i\rangle u_i,\end{equation}                          
where $u_i$ denotes the elastic displacement field. $\rho$ and
$C_{ijkl}$ are the mass density and elastic tensor. The term
$\langle h|u_i\rangle u_i/2$
represents the internal long-range action potential.\\

Using $u_i$ instead of $\varphi$ in (\ref{10}), and then inserting
Eq. (\ref{37}) in (\ref{10}), we have
\begin{equation}\label{38}\rho\ddot{u}_i+
\langle\kappa|u_i\rangle=C_{ijkl}u_{k,lj},
\end{equation}                                                                     
where the $\kappa$ is represented as
\begin{equation}\label{39}\kappa=h(\textbf{x},\textbf{y},|r_i|)+
\frac{1}{2}|r_i|\frac{\partial
h(\textbf{x},\textbf{y},|r_i|)}{\partial|r_i|},
\quad r_i=u_i(\textbf{x})-u_i(\textbf{y}).\end{equation}                                           
It is easy to see that Eq. (\ref{38}) characterizes the motion
equation of a nonlocal elasticity associated with MBCM. This theory
is an extension of the mechanically based model of nonlocal
elasticity \cite{12,13,17}. If $h$ is supposed to be independent of
$|r_i|$, then $\kappa=h$. Thus, Eq. (\ref{38}) will reduce to the
motion equation in the mechanicall based model of nonlocal
elasticity, see \cite{12,13,17}.\\

In order to find the conservation laws corresponding to Eq.
(\ref{38}), we firstly calculate the conservation current according
to Eq. (\ref{37}). Let $\varphi=u_k$. Then Eq. (\ref{33}) becomes
\begin{equation}\label{40s}J_{\alpha\gamma}=\frac{\partial
L}{\partial u_{k,\gamma}}\Phi_{k\alpha}+\{[L-\int_{\Omega}
r_ih({\bf{x}}, {\bf{y}}, |r_i|)\frac{\partial L}{\partial\langle
h|u_i\rangle}\mathrm{d}v({\bf{y}})]\delta_{\gamma\mu}-
\frac{\partial L}{\partial u_{k,\gamma}}u_{k,\mu}\}X_{\mu\alpha}.\end{equation}                    
It should be remembered that, in the convention of this paper, Latin
indices take 1, 2 and 3; while Greek indices run from 0 to 3. Since
$u_0$ is null, $\Phi_{k0}=0$. Using Eq. (\ref{37}) and (\ref{40s}),
and noticing $u_{k,0}=\dot{u}_k=\mathrm{d}u_k/\mathrm{d}t$, we have
\begin{eqnarray}\label{40}J_{\alpha\gamma,\gamma}&=&\frac{\mathrm{d}}{\mathrm{d}t}[\rho\dot{u}_k\Phi_{k\alpha}
+(L+\frac{1}{2}\int_\Omega
h(\textbf{x},\textbf{y},|r_i|)r_ku_k(t,\textbf{y})\mathrm{d}v(\textbf{y}))X_{0\alpha}-
\rho\dot{u}_k(\dot{u}_kX_{0\alpha}+u_{k,s}X_{s\alpha})]\nonumber\\
&-&[C_{ksij}u_{i,j}\Phi_{k\alpha}-(L+\frac{1}{2}\int_\Omega
h(\textbf{x},\textbf{y},|r_i|)r_ku_k(t,\textbf{y})\mathrm{d}v(\textbf{y}))X_{s\alpha}-
C_{ksij}u_{i,j}(\dot{u}_kX_{0\alpha}+u_{k,l}X_{l\alpha})]_{,s}.\end{eqnarray}                          
By Hooke's law $\sigma_{ks}=C_{ksij}u_{i,j}$, Eq. (\ref{40}) is
rewritten as
\begin{eqnarray}\label{41}J_{\alpha\gamma,\gamma}&=&\frac{\mathrm{d}}{\mathrm{d}t}[\rho\dot{u}_k\Phi_{k\alpha}
+(L+\frac{1}{2}\int_\Omega
h(\textbf{x},\textbf{y},|r_i|)r_ku_k(t,\textbf{y})\mathrm{d}v(\textbf{y}))X_{0\alpha}-
\rho\dot{u}_k(\dot{u}_kX_{0\alpha}+u_{k,s}X_{s\alpha})]\nonumber\\
&-&[\sigma_{ks}\Phi_{k\alpha}-(L+\frac{1}{2}\int_\Omega
h(\textbf{x},\textbf{y},|r_i|)r_ku_k(t,\textbf{y})\mathrm{d}v(\textbf{y}))X_{s\alpha}-
\sigma_{ks}(\dot{u}_kX_{0\alpha}+u_{k,l}X_{l\alpha})]_{,s}.\end{eqnarray}                          
In the classical elasticity, Fletcher proved the completeness of
conservation laws under the infinitesimal transformations below
\cite{28}:
\begin{equation}t\longmapsto t'=t+\epsilon(\nu t+c_0),\quad
x_i\longmapsto x'_i=x_i+\epsilon(\nu
x_i+\epsilon_{ijk}x_jb_k+c_i),\nonumber
\end{equation}
\begin{equation}\label{s1}u_i\longmapsto u'_i=u_i+\epsilon(-\nu
u_i+\epsilon_{ijk}u_jb_k+\epsilon_{ijk}x_jc_k+d_i),
\end{equation}                                                                                   
where $\epsilon$ is a infinitesimal parameter. $\nu, a_i, b_i,
c_\alpha$ and $d_i$ are arbitrary real constants. In terms of Eq.
(\ref{s1}), we turn now to investigating the concrete forms of the
conservation laws under four typical transformations.
\subsection{Case 1: $t'=t+\epsilon,\quad x'_k=x_k,\quad
u'_k=u_k$} The transformations above are equivalent to taking
$X_{00}=1, X_{s\alpha}=0, \Phi_{k\alpha}=0$ in Eq. (\ref{31}). Under
this case, Eq. (\ref{41}) reduces to
\begin{equation}\label{42}J_{\gamma,\gamma}=\frac{\mathrm{d}}{\mathrm{d}t}[L+\frac{1}{2}\int_\Omega
h(\textbf{x},\textbf{y},|r_i|)r_ku_k(t,\textbf{y})\mathrm{d}v(\textbf{y})-
\rho\dot{u}_k\dot{u}_k]+(\sigma_{ks}\dot{u}_k)_{,s}.\end{equation}                          
Substituting Eq. (\ref{42}) into (\ref{34}) yields
\begin{equation}\label{43}\int_\Omega[\frac{\mathrm{d}}{\mathrm{d}t}(L+\frac{1}{2}\int_\Omega
h(\textbf{x},\textbf{y},|r_i|)r_ku_k(t,\textbf{y})\mathrm{d}v(\textbf{y})-
\rho\dot{u}_k\dot{u}_k)+(\sigma_{ks}\dot{u}_k)_{,s}]\mathrm{d}v(\textbf{x})=0.\end{equation}                     
In terms of Eq. (\ref{2}), it is easy to verify that
\begin{eqnarray}\label{44}\int_\Omega\int_\Omega
h(\textbf{x},\textbf{y},|r_i|)r_ku_k(t,\textbf{y})\mathrm{d}v(\textbf{y})\mathrm{d}v(\textbf{x})&=&\int_\Omega
u_k(t,\textbf{y})\int_\Omega
h(\textbf{x},\textbf{y},|r_i|)r_k\mathrm{d}v(\textbf{x})\mathrm{d}v(\textbf{y})\nonumber\\&=&
\int_\Omega u_k(t,\textbf{y})\int_\Omega
\langle h|u_k\rangle\mathrm{d}v(\textbf{x})\mathrm{d}v(\textbf{y})=0.\end{eqnarray}                          
Substituting Eq. (\ref{44}) into (\ref{43}) yields
\begin{equation}\label{45}\int_\Omega[\frac{\mathrm{d}}{\mathrm{d}t}(L-
\rho\dot{u}_k\dot{u}_k)+(\sigma_{ks}\dot{u}_k)_{,s}]\mathrm{d}v(\textbf{x})=0.\end{equation}                          
By Eq. (\ref{47}) and the divergence theorem, Eq. (\ref{45}) is
rewritten as
\begin{equation}\label{46}\frac{\mathrm{d}}{\mathrm{d}t}\int_\Omega(\frac{1}{2}
\rho\dot{u}_k\dot{u}_k+\frac{1}{2}C_{ijkl}u_{i,j}u_{k,l}+
\frac{1}{2}\langle h|u_k\rangle u_k)\mathrm{d}v(\textbf{x})-
\int_{\partial\Omega}\sigma_{ks}\dot{u}_kn_s\mathrm{d}a(\textbf{x})=0.\end{equation}                          
Eq. (\ref{45}) corresponds to the conservation of energy that shows
total energy on $\Omega$, including the sum of kinetic energy,
elastic potential energy and the internal long-range action
potential energy, is equal to work done by external traction.
However, because the internal long-range interactions give rise to
energy transferring among different parts of body, the same form as
Eq. (\ref{45}) or (\ref{46}) is no longer valid for the subdomain of
$\Omega$. --This is just a intrinsic character solely processed by
the nonlocal theory.
\subsection{Case 2: $t'=t,\quad x'_k=x_k,\quad u'_k=u_k+\epsilon_k
,\quad\epsilon_1=\epsilon_2=\epsilon_3=\epsilon$} The
transformations above represent the rigid body translations. To
satisfy these transformations, we set $X_{\gamma\alpha}=0,
\Phi_{sk}=\delta_{sk}$. Thus, Eq. (\ref{41}) reduces to
\begin{equation}\label{47}J_{k\gamma,\gamma}=\frac{\mathrm{d}}{\mathrm{d}t}(\rho\dot{u}_k)
-(\sigma_{ks})_{,s}.\end{equation}                          
Substituting Eq. (\ref{47}) into (\ref{34}) leads to
\begin{equation}\label{48}\frac{\mathrm{d}}{\mathrm{d}t}\int_\Omega\rho\dot{u}_k
\mathrm{d}v(\textbf{x})-\int_{\partial\Omega}\sigma_{ks}n_s
\mathrm{d}v(\textbf{x})=0,\end{equation}                          
which is the integral representation of the conservation law of
linear momentum. Although Eq. (\ref{48}) has the same form as the
relevant conservation law in the classical elasticity, it can not be
transformed into the same differential equation due to the
nonlocality.
\subsection{Case 3: $t'=t,\quad x'_k=x_k,\quad
u'_k=u_k+\epsilon_{kij}x_i\epsilon_j,\quad
\epsilon_1=\epsilon_2=\epsilon_3=\epsilon$} The transformations
above characterize the rigid body rotations. Under this case, we
have $X_{\alpha\beta}=0, \Phi_{kj}=\epsilon_{kij}x_i$. So Eq.
(\ref{41}) leads to
\begin{equation}\label{49}J_{k\gamma,\gamma}=\frac{\mathrm{d}}{\mathrm{d}t}(\rho\epsilon_{kji}x_i\dot{u}_j)
-(\epsilon_{kji}x_i\sigma_{js})_{,s}.\end{equation}                          
Substituting Eq. (\ref{49}) into (\ref{34}) gives
\begin{equation}\label{50}\frac{\mathrm{d}}{\mathrm{d}t}\int_\Omega\rho\epsilon_{kji}x_i\dot{u}_j
\mathrm{d}v(\textbf{x})-\int_{\partial\Omega}\epsilon_{kji}x_i\sigma_{js}n_s
\mathrm{d}v(\textbf{x})=0.\end{equation}                          
Therefore, the conservation of total angular momentum on $\Omega$ is
associated with the invariance of the action functional under the
rigid body rotation.
\subsection{Case 4: $t'=t,\quad x'_k=x_k+\epsilon_k,\quad
u'_k=u_k,\quad \epsilon_1=\epsilon_2=\epsilon_3=\epsilon$} The
transformations above correspond to the coordinate translations that
are identical to setting $X_{ij}=\delta_{ij}, X_{0\alpha}=0,
X_{\alpha 0}=0, \Phi_{k\alpha}=0$. Thus, Eq. (\ref{41}) reduces to
\begin{equation}\label{51}J_{k\gamma,\gamma}=-\frac{\mathrm{d}}{\mathrm{d}t}(\rho\dot{u}_ju_{i,k})
+[(L+\frac{1}{2}\int_\Omega
h(\textbf{x},\textbf{y},|r_i|)r_ku_k(t,\textbf{y})\mathrm{d}v(\textbf{y}))\delta_{jk}
+\sigma_{ij}u_{i,k}]_{,j}.\end{equation}                                                     
Substituting Eq. (\ref{51}) into (\ref{34}) and using (\ref{44}), we
have
\begin{equation}\label{52}\frac{\mathrm{d}}{\mathrm{d}t}\int_\Omega\rho\dot{u}_iu_{i,k}
\mathrm{d}v(\textbf{x})-\int_{\partial\Omega}(Ln_k+\sigma_{ij}u_{i,k}n_j)
\mathrm{d}v(\textbf{x})=0.\end{equation}                                                     
Let $T_{jk}=L\delta_{jk}+\sigma_{ij}u_{i,k}$. $T_{jk}$ is the
so-called Eshelby tensor. In the classical elasticity, the Eshelby
tensor is independent of the integral path. However, the same
conclusion is no longer available in nonlocal elasticity. This is
because Eq. (\ref{52}) holds only on $\Omega$ as a whole. For any
$v\subset\Omega$, it ceases to be valid. Eq. (\ref{52}) represents
the conservation laws relevant to the Eshelby tensor.\\
In the formal four-dimensional space of $\hat{\Omega}$, Eq.
(\ref{46}) and (\ref{52}) can be combined in a unified expression
with the help of the energy-momentum tensor. This expression has
been given in \cite{17}.
\section{Localization of conservation laws and nonlocal residuals}
In previous sections, we have pointed out that, if the nonlocality
is concerned, the conservation laws are valid only on whole domain
occupied by body but fail on the local domain of the body.
Therefore, the localization of a conservation law is bound to result
in the nonlocal residual of conservation current appearing in the
local equilibrium equation, in which, the divergence of the
conservation current is equal to the nonlocal residual rather than
zero. Thus, by localization we can give the local equilibrium
equations respectively corresponding to Eq. (\ref{45}), (\ref{48}),
(\ref{50}) and (\ref{52}). They are formally represented as follows:
\begin{equation}\label{54}\frac{\mathrm{d}}{\mathrm{d}t}(
L-\rho\dot{u}_k\dot{u}_k)+
(\sigma_{ks}\dot{u}_k)_{,s}=\hat{E}.\end{equation}                          
\begin{equation}\label{55}\frac{\mathrm{d}}{\mathrm{d}t}(\rho\dot{u}_k)
-(\sigma_{ks})_{,s}=\hat{P}_k.\end{equation}                                                                 
\begin{equation}\label{56}\frac{\mathrm{d}}{\mathrm{d}t}(\rho\epsilon_{kji}x_i\dot{u}_j)
-(\epsilon_{kji}x_i\sigma_{js})_{,s}=\hat{M}_k.\end{equation}                          
\begin{equation}\label{57}\frac{\mathrm{d}}{\mathrm{d}t}(\rho\dot{u}_iu_{i,k})
-(L\delta_{jk}
+\sigma_{ij}u_{i,k})_{,j}=\hat{J}_k.\end{equation}                                                     
Here, $\hat{E}$ denotes the nonlocal residual, $\hat{P}_k$ the
nonlocal residual of linear momentum, $\hat{M}_k$ the nonlocal
residual of angular momentum and $\hat{J}_k$ the nonlocal residual
of the Eshelby tensor. All nonlocal residuals should be satisfy the
zero-mean codition, i.e.,
\begin{equation}\label{58}\int_\Omega(\hat{E},\hat{P}_k,\hat{M}_k,\hat{J}_k)
\mathrm{d}v(\textbf{x})=0,\end{equation}                                                 
so that the integrals of Eq. (\ref{54}), (\ref{55}), (\ref{56}) and
(\ref{57}) over $\Omega$ can return to Eq. (\ref{45}), (\ref{48}),
(\ref{50}) and (\ref{52}), respectively.

Now, we turn to determining the nonlocal residuals and the existence
of Eq. (\ref{54}) -- (63).
\subsection{Nonlocal residual of energy}
The conservation laws are firmly associated with the characters of
the Lagrangian $L(t, \textbf{x}, u_i, \dot{u_i}, u_{i,k}, \langle
h|u_i\rangle)$. In fact, we have noticed that if $L$ does not depend
explicitly upon $t$, Eq. (\ref{45}) can be determined by calculating
total derivation of $L$ with respect to $t$ in the integral sign of
the action functional $A[\varphi]$. Thus, if directly evaluating the
derivation of $L$ with respect to $t$, we have
\begin{eqnarray}\label{60}\frac{\mathrm{d}L}{\mathrm{d}t}
&=&\frac {\partial L}{\partial u_i}\dot{u_i}+\frac{\partial
L}{\partial\dot{u_i}}\ddot{u_i}+\frac{\partial L}{\partial
u_{i,k}}\dot{u_i}_{,k}+\frac{\partial L}{\partial \langle
h|u_i\rangle}\frac{\mathrm{d}\langle
h|u_i\rangle}{\mathrm{d}t}\nonumber\\&=&[\frac{\partial L}{\partial
u_i}-\frac{\mathrm{d}}{\mathrm{d}t}(\frac {\partial
L}{\partial\dot{u_i}})-(\frac{\partial L}{\partial
u_{i,k}})_{,k}]\dot{u_i}+\frac{\mathrm{d}}{\mathrm{d}t}(\frac
{\partial L}{\partial\dot{u_i}}\dot{u_i})+(\frac{\partial
L}{\partial u_{i,k}}\dot{u_i})_{,k}+\frac{\partial L}{\partial
\langle h|u_i\rangle}\langle g|\dot{u_i}\rangle.\end{eqnarray}                                            
Applying Eq. (\ref{10}) to (\ref{60}) gives
\begin{equation}\label{61}\frac{\mathrm{d}L}{\mathrm{d}t}
=\frac{\mathrm{d}}{\mathrm{d}t}(\frac {\partial
L}{\partial\dot{u_i}}\dot{u_i})+(\frac{\partial L}{\partial
u_{i,k}}\dot{u_i})_{,k}+\frac{\partial L}{\partial \langle
h|u_i\rangle}\langle g|\dot{u_i}\rangle-\langle g|\frac{\partial
L}{\partial\langle h|u_i\rangle}\rangle\dot{u_i}.\end{equation}                                            
Eq. (\ref{61}) can be rewritten as
\begin{equation}\label{62}\frac{\mathrm{d}}{\mathrm{d}t}
(L-\frac {\partial L}{\partial\dot{u_i}}\dot{u_i})-(\frac{\partial
L}{\partial u_{i,k}}\dot{u_i})_{,k}=\frac{\partial L}{\partial
\langle h|u_i\rangle}\langle g|\dot{u_i}\rangle-\langle
g|\frac{\partial
L}{\partial\langle h|u_i\rangle}\rangle\dot{u_i}.\end{equation}                                            
By using Eq. (\ref{37}), Eq. (\ref{62}) leads to
\begin{equation}\label{63}\frac{\mathrm{d}}{\mathrm{d}t}(
L-\rho\dot{u}_i\dot{u}_i)+
(\sigma_{ik}\dot{u}_i)_k=\dot{u}_i(t,\textbf{x})\int_\Omega
g(\textbf{x}, \textbf{y}, |r_i|)u_i(t,
\textbf{y})\mathrm{d}v(\textbf{x})-u_i(t,\textbf{x})\int_\Omega
g(\textbf{x},\textbf{y},|r_i|)\dot{u}_i(t, \textbf{y})dv(\textbf{y}).\end{equation}                              
Compared Eq. (\ref{63}) with (\ref{54}), we have
\begin{equation}\label{64}\hat{E}=\dot{u}_i(t,\textbf{x})\int_\Omega g(\textbf{x}, \textbf{y},
|r_i|)u_i(t,\textbf{y})\mathrm{d}v(\textbf{x})-u_i(t,\textbf{x})\int_\Omega
g(\textbf{x},\textbf{y},|r_i|)\dot{u}_i(t, \textbf{y})dv(\textbf{y}).\end{equation}                                
So far, $\hat{E}$ has been determined. Clearly, it satisfies Eq.
(\ref{58}).
\subsection{Nonlocal residuals of linear and angular momentum}
Eq. (\ref{55}) characterizes the equilibrium of linear momentum. So
it is identical to Eq. (\ref{38}). Due to this fact, it is easy to
see that $\hat{P}_k=-\langle\kappa|u_k\rangle$, i.e.,
\begin{equation}\label{65}\hat{P}_k=\int_\Omega
\kappa(\textbf{x},\textbf{y},|r_i|)u_k(t,
\textbf{y})dv(\textbf{y})-u_k(t,\textbf{x})\int_\Omega\kappa(\textbf{x},
\textbf{y},
|r_i|)\mathrm{d}v(\textbf{x}).\end{equation}                                            
$\hat{P}_k$ can be used to represent $\hat{M}_k$. In order to verify
this point, we firstly rewritten Eq. (\ref{56}) as follows:
\begin{eqnarray}\label{66}\hat{M}_k&=&\frac{\mathrm{d}}{\mathrm{d}t}(\rho\epsilon_{kji}x_i\dot{u}_j)
-(\epsilon_{kji}x_i\sigma_{js})_{,s}\nonumber\\&=&\epsilon_{kji}x_i[\frac{\mathrm{d}}{\mathrm{d}t}(\rho\dot{u}_j)
-(\sigma_{js})_{,s}].\end{eqnarray}                          
Inserting Eq. (\ref{55}) in (\ref{66}) gives
\begin{equation}\label{67}\hat{M}_k=\epsilon_{kji}x_i\hat{P}_j.\end{equation}            
In a general case, it is easy to see that
$\epsilon_{kji}x_i\hat{P}_j$ always violates the zero-mean
condition. As a result, $\hat{M}_k$ given by Eq. (\ref{67}) does not
satisfy Eq. (\ref{58}). On the other hand, $\hat{M}_k$ should follow
Eq. (\ref{58}) so as to ensure the integrals of Eq. (\ref{56}) over
$\Omega$ returning to Eq. (\ref{50}). The paradox caused by Eq.
(\ref{67}) shows that it is impossible, under a general case, to
transform Eq. (\ref{50}) into the form of Eq. (\ref{56}) by
localization. Eq. (\ref{56}) is inaccessible.\\

However, if the nonlocal kernel takes the form of the central pair
potential, $\epsilon_{kji}x_i\hat{P}_j$ can be consistent with the
zero-mean condition. Under this circumstance, $\langle h|u_k\rangle$
is simplified into $\langle h|x_k\rangle$, and $\hat{P}_k$ can be
represented as
\begin{equation}\label{68}\hat{P}_k=\int_\Omega
\kappa(\textbf{x},\textbf{y},|r_i|)y_kdv(\textbf{y})-x_k\int_\Omega\kappa(\textbf{x},
\textbf{y},
|r_i|)\mathrm{d}v(\textbf{x}).\end{equation}                                            
Substituting Eq. (\ref{68}) into (\ref{67}) yields
\begin{equation}\label{69}\hat{M}_k=\epsilon_{kji}x_i\int_\Omega
\kappa(\textbf{x},\textbf{y},|r_i|)y_i\mathrm{d}v(\textbf{y}),\end{equation}                              
which is equivalent to the expression below:
\begin{equation}\label{70}\hat{\textbf{M}}=\textbf{x}\times\int_\Omega
\kappa(\textbf{x},\textbf{y},|r_i|)\textbf{y}\mathrm{d}v(\textbf{y}).\end{equation}                              
Since both \textbf{x} and \textbf{y} are the dummy variables of
integration in the below, we have
\begin{eqnarray}\label{71}\int_\Omega\hat{\textbf{M}}\mathrm{d}v(\textbf{x})&=&\int_\Omega\int_\Omega
\kappa(\textbf{x},\textbf{y},|r_i|)\textbf{x}\times\textbf{y}\mathrm{d}v(\textbf{y})\mathrm{d}v(\textbf{x})
\nonumber\\&=&\int_\Omega\int_\Omega
\kappa(\textbf{y},\textbf{x},|r_i|)\textbf{y}\times\textbf{x}\mathrm{d}v(\textbf{x})\mathrm{d}v(\textbf{y})
\nonumber\\&=&-\int_\Omega\int_\Omega
\kappa(\textbf{x},\textbf{y},|r_i|)\textbf{x}\times\textbf{y}\mathrm{d}v(\textbf{y})\mathrm{d}v(\textbf{x})
=-\int_\Omega\hat{\textbf{M}}\mathrm{d}v(\textbf{x}).\end{eqnarray}                              
Eq. (\ref{71}) leads to
\begin{equation}\label{72}\int_\Omega\hat{\textbf{M}}\mathrm{d}v(\textbf{x})=0,
\quad \mathrm{i.e.},\quad\int_\Omega\hat{M}_k\mathrm{d}v(\textbf{x})=0.\end{equation}                            
So far, we have demonstrated that $\hat{M}_k$ satisfies the
zero-mean condition when the long-range body force is governed by a
central pair potential.
\subsection{Nonlocal residual of the Eshelby tensor}
The conservation law relevant to the Eshelby tensor attributes to
the invariance of the action functional under the coordinate
translations. Therefore, both the Lagrangian $L$ and the nonlocal
kernel $h$ are bound to be explicitly independent of $x_k$. As thus,
we have
\begin{eqnarray}\label{73}L_{,k}&=&\frac{\partial L}{\partial
u_i}u_{i,k}+\frac{\partial L}{\partial
\dot{u}_i}\dot{u}_{i,k}+\frac{\partial L}{\partial
u_{i,j}}u_{i,jk}+\frac{\partial L}{\partial \langle
h|u_i\rangle}\langle h|u_i\rangle_{,k}\nonumber\\&=&[\frac{\partial
L}{\partial u_i}-\frac{\mathrm{d}}{\mathrm{d}t}(\frac{\partial
L}{\partial \dot{u}_i})-(\frac{\partial L}{\partial
u_{i,j}})_{,j}]u_{i,k}+\frac{\mathrm{d}}{\mathrm{d}t}(\frac{\partial
L}{\partial \dot{u}_i}u_{i,k})+(\frac{\partial L}{\partial
u_{i,j}}u_{i,k})_{,j}+\frac{\partial L}{\partial
\langle h|u_i\rangle}\langle h|u_i\rangle_{,k}.\end{eqnarray}                         
Applying Eq. (\ref{10}) to (\ref{73}) leads to
\begin{equation}\label{74}L_{,k}=\frac{\mathrm{d}}{\mathrm{d}t}(\frac{\partial
L}{\partial \dot{u}_i}u_{i,k})+(\frac{\partial L}{\partial
u_{i,j}}u_{i,k})_{,j}+\frac{\partial L}{\partial \langle
h|u_i\rangle}\langle h|u_i\rangle_{,k}-
\langle g|\frac{\partial L}{\partial \langle h|u_i\rangle}\rangle u_{i,k}.\end{equation}                         
Substituting Eq. (\ref{37}) into (\ref{74}) yields
\begin{equation}\label{75}L_{,k}=\frac{\mathrm{d}}{\mathrm{d}t}(\rho\dot{u}_iu_{i,k})-(\sigma_{ij}u_{i,k})_{,j}
+u_i\langle h|u_i\rangle_{,k}-\langle g|u_i\rangle u_{i,k}.\end{equation}                         
Eq. (\ref{75}) is rewritten as
\begin{equation}\label{76}\frac{\mathrm{d}}{\mathrm{d}t}(\rho\dot{u}_iu_{i,k})-(L\delta_{kj}+\sigma_{ij}u_{i,k})_{,j}
=\langle g|u_i\rangle u_{i,k}-u_i\langle h|u_i\rangle_{,k}.\end{equation}                         
Comparison between Eq. (\ref{75}) and Eq. (\ref{57}) will gives
\begin{equation}\label{77}\hat{J}_k=\langle g|u_i\rangle u_{i,k}-
u_i\langle h|u_i\rangle_{,k}.\end{equation}                                                      
Clearly, $\hat{J}_k$ determined by Eq. (\ref{77}) fails to agree
with the zero-mean condition. -- This means that the nonlocal
residual of the Eshelby tensor does not exist. Therefore, Eq.
(\ref{76}) is not a local equilibrium equation corresponding to Eq.
(\ref{52}). In other word, Eq. (\ref{52}) can not be transformed
into a local form similar to Eq. (\ref{57}). The conservation law
relevant to the Eshelby tensor exists only in the form of
integration.
\section{Conclusions}
\label{sec:4} In this paper, we extend the definition of the
nonlocal argument by introducing a nonlinear nonlocal kernel
depending not only on the spatial coordinates but also on the field
variables. The extended definition retains the zero mean character
of the nonlocal argument. It is this character to distinguish the
nonlocal Lagrangian formulation developed in this paper from other
nonlocal variational theories.\\

On the basis of the extended nonlocal argument, a Lagrangian
formulation with nonlocality is established. The nonlocal
Euler-Lagrange equation is derived from the Hamilton's principle.
Accompanied with this equation, the nonlocal traction appears in the
form of the nonlocal residual satisfying the zero mean condition
automatically. -- This is an obvious difference between the new
theory and the other theories. Physically, the nonlocal traction
represents the long-range interactions within body. Therefore, the
zero-mean character of the nonlocal traction is just a embodiment of
the action and
reaction law.\\

The Noether's theorem is extended to the Lagrangian formulation with
nonlocality. The result shows that conservation law exists only in
the form of the integral over the whole domain occupied by body. The
local equilibrium equation derived from the localization of the
conservation law is equal to the nonlocal residual of the
conservation current, provided it exists, rather than zero like the
case in the variational theories without nonlocality. In physics,
the presence of the nonlocal residual attributes to the conservation
current transferring, caused by the internal long-range
interactions, among different parts within body.\\

A Lagrangian including the nonlocal argument is advanced in the
quadratic form. The motion equation derived from this Lagrangian is
consistent with the nonlocal elasticity associated with MBCM. In
this theoretical framwork, we use the extended Noether's theorem to
determine the conservation laws relevant to energy, linear momentum,
angular momentum and the Eshelby tensor. The localization of these
conservation laws are discussed in detail. We demonstrate that local
equilibrium equation of energy and of linear momentum can be
respectively derived from the corresponding conservation laws by
localization, but the conservation law relevant to the Eshelby
tensor can not be transformed into a local form by localization. So
no local equilibrium equation relevant to the Eshelby tensor exists
in the nonlocal elasticity associated with MBCM. The nonlocal
residual of energy and nonlocal residual of linear momentum have
been determined, respectively. They are consistent in mathematical
form with the
representation of nonlocal residual given in \cite{27}.\\

In general, there is no local equilibrium equation corresponding to
the conservation law of angular momentum. However, if the nonlocal
kernel takes the form of the central pair potential, the local
equilibrium equation of angular momentum will occur in the nonlocal
elasticity associated with MBCM. Under this case, the internal
long-range interaction manifests itself as a central force
field.\\

We therefore conclude that, in the nonlocal elasticity associated
with MBCM, not every conservation law corresponds to a local
equilibrium equation. Only when the nonlocal residual of
conservation current exists, can a conservation law be transformed
into a local equilibrium equation by localization. The results in
this paper imply that he existence of the local equilibrium equation
is to some degree influenced by the nonlocal kernel, However, that
is a problem needing further exploration.
\section*{Acknowledgements}
The support of the National Nature Science Foundation of China
through the Grant No. GAA12012 is gratefully acknowledged. .

\end{document}